\documentclass[fleqn,12pt]{SelfArx} 

\definecolor{color1}{RGB}{0,90,30} 
\definecolor{color2}{RGB}{20,20,0} 

\newlength{\tocsep} 
\usepackage{lipsum} 
\usepackage{natbib}
\usepackage{hyperref}
\usepackage[symbol]{footmisc}

\JournalInfo{Astro2020 APC White Paper: Optical and Infrared Observations from the Ground} 
\Archive{10 July 2019} 

\PaperTitle{High-resolution Infrared Spectrograph for Exoplanet Characterization with the Keck and Thirty Meter Telescopes} 

\ShortTitle{High-resolution Infrared Spectrograph for Exoplanet Characterization with Keck and TMT}

\Authors{\normalsize Dimitri Mawet\textsuperscript{1,2}**, Michael Fitzgerald\textsuperscript{3}, Quinn Konopacky\textsuperscript{4}, Charles Beichman\textsuperscript{1,2,5}, Nemanja Jovanovic\textsuperscript{1}, Richard Dekany\textsuperscript{1}, David Hover\textsuperscript{1}, Eric Chisholm\textsuperscript{6}, David Ciardi\textsuperscript{5}, \'Etienne Artigau\textsuperscript{7}, Ravinder Banyal\textsuperscript{8}, Thomas Beatty\textsuperscript{9}, Bj\"orn Benneke\textsuperscript{7}, Geoffrey A. Blake\textsuperscript{1}, Adam Burgasser\textsuperscript{4}, Gabriela Canalizo\textsuperscript{10}, Guo Chen\textsuperscript{11}, Tuan Do\textsuperscript{3}, Greg Doppmann\textsuperscript{12}, Ren\'e Doyon\textsuperscript{7}, Courtney Dressing\textsuperscript{13}, Min Fang\textsuperscript{14}, Thomas Greene\textsuperscript{15}, Lynne Hillenbrand\textsuperscript{1}, Andrew Howard\textsuperscript{1}, Stephen Kane\textsuperscript{10}, Tiffany Kataria\textsuperscript{2}, Eliza Kempton\textsuperscript{16}, Heather Knutson\textsuperscript{1}, Takayuki Kotani\textsuperscript{17}, David Lafreni\`ere\textsuperscript{7}, Chao Liu\textsuperscript{18}, Shogo Nishiyama\textsuperscript{19}, Gajendra Pandey\textsuperscript{8}, Peter Plavchan\textsuperscript{20}, Lisa Prato\textsuperscript{21}, S.P. Rajaguru\textsuperscript{8}, Paul Robertson\textsuperscript{22}, Colette Salyk\textsuperscript{23}, Bun'ei Sato\textsuperscript{24}, Everett Schlawin\textsuperscript{9}, Sujan Sengupta\textsuperscript{8}, Thirupathi Sivarani\textsuperscript{8}, Warren Skidmore\textsuperscript{6}, Motohide Tamura\textsuperscript{25}, Hiroshi Terada\textsuperscript{6, 26}, Gautam Vasisht\textsuperscript{2}, Ji Wang\textsuperscript{27}, Hui Zhang\textsuperscript{28}}

\affiliation{\textsuperscript{1}\textit{California Institute of Technology, CA, USA.}
\textsuperscript{2}\textit{Jet Propulsion Laboratory, CA, USA.}
\textsuperscript{3}\textit{University of California, Los Angeles, CA, USA.}
\textsuperscript{4}\textit{University of California, San Diego, CA, USA.}
\textsuperscript{5}\textit{NASA Exoplanet Science Center, CA, USA.}
\textsuperscript{6}\textit{Thirty Meter Telescope, CA, USA.}
\textsuperscript{7}\textit{Universit\'e de Montr\'eal, QC, Canada.}
\textsuperscript{8}\textit{Indian Institute of Astrophysics, Bangalore, India.}
\textsuperscript{9}\textit{University of Arizona, AZ, USA.}
\textsuperscript{10}\textit{University of California, Riverside, CA, USA.}
\textsuperscript{11}\textit{Purple Mountain Observatory, Nanjing, China.}
\textsuperscript{12}\textit{W.M. Keck Observatory, HI, USA.}
\textsuperscript{13}\textit{University of California, Berkeley, CA, USA.}
\textsuperscript{14}\textit{Xiamen University, Fujian China.}
\textsuperscript{15}\textit{NASA Ames Research Center, CA, USA.}
\textsuperscript{16}\textit{University of Maryland, MD, USA.}
\textsuperscript{17}\textit{Astrobiology Center, National Institutes of Natural Sciences, Japan.}
\textsuperscript{18}\textit{National Astronomical Observatories, Chinese Academy of Sciences, Beijing, China.}
\textsuperscript{19}\textit{Miyagi University of Education, Sendai, Japan.}
\textsuperscript{20}\textit{George Mason University, VA, USA.}
\textsuperscript{21}\textit{Lowell Observatory, AZ, USA.}
\textsuperscript{22}\textit{University of California, Irvine, CA, USA.}
\textsuperscript{23}\textit{Vassar College, NY, USA.}
\textsuperscript{24}\textit{Tokyo Institute of Technology, Tokyo, Japan.}
\textsuperscript{25}\textit{University of Tokyo, Tokyo, Japan.}
\textsuperscript{26}\textit{National Astronomical Observatory of Japan, Tokyo, Japan.}
\textsuperscript{27}\textit{Ohio State University, OH, USA.}
\textsuperscript{28}\textit{Nanjing University, Nanjing, China.}
}

\affiliation{** \textbf{E-mail, phone}: dmawet@astro.caltech.edu, 626-395-1452} 
\newcommand{\keywordname}{Keywords} 

\Abstract{HISPEC (High-resolution Infrared Spectrograph for Exoplanet Characterization) is a proposed diffraction-limited spectrograph for the W.M. Keck Observatory, and a pathfinder for the MODHIS facility project (Multi-Object Diffraction-limited High-resolution Infrared Spectrograph) on the Thirty Meter Telescope. HISPEC/MODHIS builds on diffraction-limited spectrograph designs such as Palomar-PARVI and LBT-iLocator, both of which rely on adaptively corrected single-mode fiber feeds. Seeing-limited high-resolution spectrographs, by virtue of the conservation of beam etendue, grow in volume following a $D^3$ power law ($D$ is the telescope diameter), and are subject to daunting challenges associated with their large size (e.g. mechanical and thermal stability). Diffraction-limited spectrographs fed by single mode fibers are decoupled from the telescope input, and are orders of magnitude more compact and have intrinsically more stable line spread functions. On the flip side, their efficiency is directly proportional to the performance of the adaptive optics (AO) system. AO technologies have matured rapidly over the past two decades, becoming mainstream on current large ground-based telescopes and baselined for future extremely large telescopes. HISPEC/MODHIS will take R$>$100,000 spectra of a few objects in a 10" field-of-view sampled at the diffraction limit (10 mas scale), simultaneously from 0.95 to 2.4 $\mu$m (y band to K band). The scientific scope ranges from exoplanet infrared precision radial velocities, transit and close-in exoplanet spectroscopy (atmospheric composition and dynamics, RM effect), spectroscopy of directly imaged planets (atmospheric composition, spin measurements, Doppler imaging), brown dwarf characterization, stellar physics/chemistry, proto-planetary disk kinematics/composition, Solar system (e.g. comets), extragalactic science, and cosmology. HISPEC/MODHIS features a compact and cost-effective design optimized to fully exploit the existing Keck-AO and future TMT-NFIRAOS infrastructures and boost the scientific reach of both Keck Observatory and TMT soon after first light.}

\begin{document}

\flushbottom 

\maketitle 
\pagenumbering{arabic}
\clearpage


\section{Key Science Goals and Objectives}

HISPEC/MODHIS will offer a powerful new window into a range of
science topics, from exoplanets to distant galaxies. Moreover, because of its compact, diffraction-limited design, MODHIS is targeted as a first light instrument for TMT which will provide a high spectral resolution capability that would not otherwise be possible for many years.

\begin{table*}[h!]
\centering
 \begin{tabular}{|l|l|} 
 \hline
 Spectral resolution & $>$100,000 \\ 
 \hline
 Wavelength coverage & 0.95-2.4 $\mu$m (yJHK) simultaneous\\
 \hline
 Multiplexing &1-9 channels, including object, sky, calibration \\
 \hline
 Angular resolution at y band & 7 mas (TMT) - 20 mas (Keck) \\
 \hline
 Angular resolution at K band & 15 mas (TMT) - 44 mas (Keck) \\
 \hline
 Field of regard & 10" patrol diameter\\
 \hline
 High contrast capabilities & $10^{-3}$ raw contrast at the diffraction limit $\lambda/D$\\
 \hline
 Point-source limiting mag (1 hr, S/N=10 per sp. ch.) &17 (Keck) -- 19 (TMT) mag\\
 \hline
 Calibration & Laser Frequency Comb, Etalon, Gas cells\\
 \hline
 Instrumental stability &30 cm/s\\
 \hline
 \end{tabular}
\caption{HISPEC/MODHIS specifications.}
\label{table:specs}
\end{table*}


\subsection{High Spectral Resolution Characterization of Exoplanets}
Using high-dispersion spectroscopy as a way of spectral filtering has been successfully demonstrated in a few studies.  The improved resolution, sensitivity, and Line Spread Function (LSF) stability provided by the Single Mode Fiber (SMF) fed diffraction-limited spectrograph designs will help mitigate systematic noise introduced when isolating signatures of the planet from those of the star and terrestrial atmosphere.

\paragraph{Characterization of Close-In Planets.}
For transiting and non-transiting hot Jupiters and warm Neptunes, high-resolution transmission spectroscopy has been used to detect molecular gas \citep{snellen2010, birkby2013, dekok2014} and to study day to night side wind velocity \citep{snellen2010}, providing an ultimate test for 3D exoplanet atmosphere models \citep{miller2012}.  For planets detected with the RV method, spectral lines owing to the planet and star may be separated via their radial velocities ($>$50 km/s).  The RV of a planet can thus be measured to break the degeneracy between the true planet mass and orbital inclination intrinsic to RV detection \citep{brogi2012,brogi2013,brogi2014,lockwood2014}.  HISPEC will provide the sensitivity, spectral resolution, and spectral coverage necessary for follow-up opportunities in the TESS era. MODHIS on TMT will easily push the sensitivity boundaries to sub-Neptunes and super-Earths \citep{Dragomir2019}\footnotemark[2], including the possibility of detecting Earth-like biosignatures on rocky exoplanets around nearby stars \citep{Lopez-Morales2019}\footnotemark[2].
\footnotetext[2]{Denotes an Astro2020 Science White Paper}

\paragraph{Characterization of Long-Period Imaged Planets.}

Coupling a high-resolution spectrograph with a high-contrast imaging instrument will enable the direct characterization of exoplanet atmospheres (\citealt{snellen2015,bowler2019}\footnotemark[2]).  In this scheme, the AO system serves as a spatial filter, separating the light from the star and the planet, and the spectrograph serves as the spectral filter, which differentiates between features in the stellar and planetary spectra \citep{hdc1,hdc2}.  High-resolution spectroscopy has three game changing benefits: 1- Detailed species-by-species molecular characterization, abundance ratios such as [C/O]. 2- Doppler measurements of the planet's spin \citep{snellen2014}, orbital velocity, plus mapping of atmospheric and/or surface features \citep{crossfield2014}. 3- Improved detection capability \citep{hdc1,hdc2} by side-stepping speckle noise calibration issues affecting low-resolution spectroscopic data from current integral field spectrographs such as SPHERE \citep{claudi2008} and GPI \citep{macintosh2007}. HISPEC on Keck will address the detailed spectroscopic characterization of young giant planets, but using the same technique with MODHIS on TMT will enable the direct detection and characterization of Neptune-size and possibly Earth-size exoplanets \citep{Wang2019}\footnotemark[2], although the latter case will generally require an extreme AO front-end instrument such as PSI (see white paper from Fitzgerald et al.~2019).

\subsection{Detection of Exoplanets at or Within the Diffraction Limit}

Fiber nulling as a means to detect and characterize exoplanets and circumstellar disks at or within the diffraction limit was first introduced and demonstrated by \citet{Haguenauer2006}. This concept is largely based on the Bracewell nulling interferometer \citep{Bracewell1979}. The first fiber nuller was demonstrated on sky at Palomar observatory  by \citet{Hanot2011}. \citet{Ruane2018b} recently introduced the concept of vortex fiber nulling (VFN), which circumvents the need of a rotating baseline and greatly simplifies the design and operation of the fiber nuller. The VFN concept was demonstrated in the laboratory by \citet{Echeverri2019}, and will be the subject of an on-sky science demonstration at Keck observatory in 2020. A VFN mode is currently baselined for HISPEC/MODHIS. The combination of VFN starlight suppression at the $10^{-3}$ raw contrast level (limited by AO residuals and finite stellar size) and high-resolution spectroscopy will enable the detection and high-resolution spectroscopic characterization of planets at or within the diffraction limit (down to $\simeq 10$ mas on TMT). Using the latest giant planet occurrence rates from \citet{Nielsen2019}, we predict the discovery and simultaneous characterization of dozens of new young giant planets in nearby young associations and star forming regions with both Keck-HISPEC, and even more with TMT-MODHIS. Benefiting from the giant aperture and angular resolution of TMT, \citet{Ruane2018b} predicts the detection of Ross 128 b in reflected light in 30 hours. Less challenging configurations, which include giant planet, mini-Neptunes and super-Earths will also be accessible. The most challenging cases of temperate Earth-size planets in less favorable configurations than Ross 128 b (e.g. more distant) will require PSI (Fitzgerald et al.~2019).

\begin{figure}[!t]
        \centering
        \includegraphics[width=0.5\textwidth]{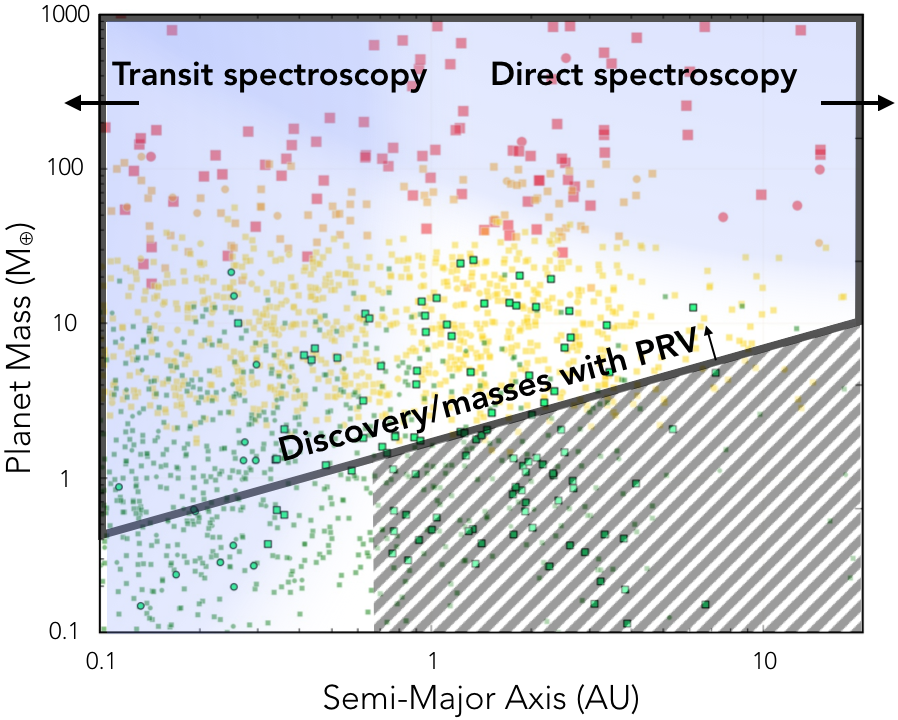}
        \caption{HISPEC/MODHIS potential exoplanet discovery and characterization space in a Mass (Earth masses) vs Separation (SMA in AU) diagram. The various markers denote simulated planet populations within 27 pc using the occurrence rates from \citet{Kopparapu2018} extrapolated up to a semi-major axis of 30 AU using an exponential cutoff. Various cutoffs in magnitude were applied to reflect TMT photon noise limits in 10 hours. The marker size is proportional to the planet size: red for giant planets (radius $>$  R$_\oplus$); orange for Neptunes (6 $>$ R$_\oplus$ $>$ 3.5); yellow for mini-Neptunes (3.5 $>$ R$_\oplus$ $>$ 1.75); dark green for super-Earth and Earth-size planets (1.75 $>$ R$_\oplus$ $>$ 0.5); and light green for temperate ([$0.7\sqrt(L/L_\odot)$, $1.5\sqrt(L/L_\odot)$] AU) super-Earth and Earth-size planets. The round markers are for planets around cool stars (T$_{eff}$ $<$ 4000 K), while the square markers denote planets around warmer stars (T$_{eff}$ $>$ 4000 K).}
        \label{fig:modhis_phasespace}
\end{figure}

\subsection{Exoplanet detection and masses with PRV}
Though we know of a multitude of planetary systems from
transit missions, planetary masses are essential for
constraining their properties, including densities and
atmospheric properties \citep{Batalha2017}.  Thus, the need for new precision radial velocity (PRV) instrumentation extends not only to the current 8-10 meter
telescope facilities, but beyond into the ELT era, where large
collecting areas will allow for the collection of high SNR
spectra in short enough time to obtain information about
stellar activity \citep{Ciardi2019}\footnotemark[2].




\subsubsection{Stellar Jitter and the NIR Advantage} \label{sec:stellarjitter}
Doppler monitoring at NIR wavelengths offers two enormous assists to the goal of measuring the masses of exoplanets orbiting young and/or cool host stars.  First, both cool stars and the majority of young stars are brighter in the infrared than in the optical yielding higher SNR at longer wavelengths.  Second, there are theoretical and observational reasons to expect that RV noise is reduced in the NIR \citep{plavchan2015}.  Recent studies of M stars with CARMENES show this tendency clearly in the y-band and future work is likely to extend this trend further into the NIR \citep{talor2018}.

Studies of very young stars have established that the amplitude of RV variability is a factor of 2-4 lower in the NIR, where spot-to-photosphere temperature contrasts are lower \citep{Prato2008,mahmud2011,Crockett2012,carleo2018}. \citet{robertson2016} found that the K I doublet (7665 and 7699~\AA) is a good indicator of stellar activity levels, leading us to investigate the utility of the $J$-band K I doublet at 12,400~\AA\@ and 12,500~\AA\@ for assessing correlations between chromospheric activity and RV variability.  HISPEC/MODHIS's high spectral resolution will also aid in characterizing Zeeman splitting as an indicator of stellar jitter \citep{moutou2017}.

\subsubsection{NIR PRV science cases}

The  radius  and ephemeris of transiting planets are known \textit{a priori} and the expected RV amplitude can also be predicted to within a factor of a few.  Further,  the rotation periods and characteristic lifetimes of the surface features of host stars can often be determined from the  high precision time series photometry, critical information  for the modeling of time series RV measurements to remove stellar activity  \citep{aigrain2012, haywood2014, dai2017}.


\paragraph{Young Planets.} Planets transiting young host stars have now been discovered from K2 observations of young moving groups ($\sim$50-90 Myr) \citep{david2016,david2018} and open clusters ($\sim$600-800 Myr) \citep{mann2016,mann2017,ciardi2018},  with doubtless many more to be discovered by TESS.  Among these are "Hot Jupiters" with large predicted Doppler semi-amplitudes ($\sim$10-100 m/s). With the expected reduction of  stellar jitter in the NIR, the intrinsic stellar variability at HISPEC/MODHIS wavelengths will be  comparable to or at worst a factor of 2 larger than the reflex motion due to the planets. By combining transit radii and PRV masses it will be possible to determine the density of these still contracting planets and thus to open a new era in the study of the formation of gas giant planets.


\paragraph{Transiting Habitable Zone (HZ) Planets. \label{RM}}  The transiting HZ planets most favorable for follow-up spectroscopy will orbit cool stars, e.g. Trappist-1 with 7 orbiting planets \citep{grimm2018}, and the nearest of these will be prime targets for spectroscopic studies of their atmospheres. TESS radii combined with HISPEC/MODHIS masses will yield planet density to determine whether a planet is rocky, icy, or gaseous, as well as the surface gravity which is needed for the interpretation of transit spectroscopy.  
    
\paragraph{Orbital Architectures for Small Transiting Planets.}  The angle between the stellar spin axis and the angular momentum vector of the planet orbit, or orbital obliquity, provides rich information on the history of planet formation and evolution \citep{winn2015}.  Obliquity is measured via the Rossiter-McLaughlin (RM) effect \citep{rossiter1924,mclaughlin1924} by monitoring the change in radial velocity for the duration of a transit, which usually lasts for only a few hours, or less than half an hour for planets in the habitable zones of the coolest M dwarfs. HISPEC/MODHIS will provide the sensitivity, spectral resolution and temporal resolution needed to expand  the current sample of small planets whose RM effect has been measured to date \citep{hirano2012, albrecht2013, huber2013, sanchis2015}. With MODHIS on TMT, hundreds of small planets will become accessible resulting in unique insights into how these planets form and migrate \citep{Johnson2019}\footnotemark[2].


    
\paragraph{Surveying the Coolest Mature Stars.}  
HISPEC / MODHIS will have an important niche for the coolest stars requiring the highest sensitivity \citep{reiners2018} and for binary systems requiring AO to separate the components.  For example, there are over 600 stars with spectral types between M9 and L9 with $\delta>-30^\circ$ and H $<$ 14 mag \citep{best2018}, which Keck-HISPEC could study with \textit{instrument-limited} precision of $<$ 30 cm/s.  With just a few observations per object, HISPEC could screen these systems for RV variations and then look for planets where the problem of rapid rotation ($>$ 20 km/s) would be at least partially offset by the large signals expected for Uranus-mass objects (e.g. 20 m/s for a Uranus orbiting a 50 M$_{Jup}$ brown dwarf in a 30 day orbit). MODHIS on TMT will provide improved angular resolution and sensitivity so closer separation and fainter binaries can be addressed.

\subsection{Stellar and Planetary System Formation}
\paragraph{Proto-planetary Disks.} 


The evolution, and ultimately the dispersal of proto-planetary (PP) disks, holds the key to many open questions related to planet formation.  High-resolution infrared spectroscopy is instrumental in understanding PP disk gas dynamics (e.g. the study of the emission and absorption of CO ro-vibrational lines).  Spatially resolved high-resolution spectro-astrometry (at mas scales) of the molecular gas allows one to measure its distribution in space and velocity, and to correlate these measurements with disk geometries and accretion activity. Infrared spectroscopy also enables detailed characterization of the molecular composition of young disks, including organic molecules \citep{JangCondell2019}\footnotemark[2].
    
\paragraph{Detecting/Characterizing Young Forming Planets.}

Answering the key question "How do giant planets form?" cannot be achieved by only observing older, dynamically evolved systems.  Imaging extrasolar giant planets near the epoch of their formation is much easier, due to their higher luminosities.  This requires the full angular resolution of a large telescope such as Keck or TMT, in order to separate giant planets at solar system scales from their host stars \citep{sallum2019}\footnotemark[2].  Most of the luminosity of these forming planets is expected to be emitted at NIR wavelengths \citep{zhu2015} from a circumplanetary disk.  It has so far proved difficult to uniquely separate circumplanetary emission from circumstellar disk emission, as this requires resolving line emission at high spectral dispersion.  The kinematic signatures of these warm disks will occur in the range of 3-30 km/s (similar to the velocities of Jupiter's moons), with strong line emission from CO and H$_2$O in the bandpass of HISPEC.  These data might even lead to measurements of the  dynamical masses of these forming exoplanets, in much the same way as one can derive masses of T Tauri stars from the velocity curves of circumstellar disks.  Spatially resolved, high-resolution spectroscopy of young giant planets will also enable searches for accretion signatures, if the planet is still growing.  Finally, Doppler monitoring of directly-imaged planets of any age could lead to discoveries of exomoons \citep{vanderburg2018}.
    
\subsection{Physics of Very Low Mass Objects}

HISPEC/MODHIS have the potential to revolutionize our understanding of the detailed properties of brown dwarfs, the cousins of gas giant planets.  The simultaneous wavelength coverage provided by these instruments will allow for detailed weather monitoring on L and T dwarfs, while the high spectral resolution offers the opportunity for Doppler imaging of their surfaces \citep{crossfield2014}.  The strength of their magnetic fields can be probed with high resolution spectroscopy, and kinematic information gleaned from measurements of the RVs of the lowest mass cluster members \citep{burgasser2019}\footnotemark[2].

Furthermore, PRV measurements of binary systems, enabled by the high spatial resolution of the Laser Guide Star (LGS) AO feed, will advance our understanding of the physics of very low mass objects in a number of ways:  establishing highly precise 3D orbits of short period sub-stellar objects to establish mass benchmarks for testing evolutionary/spectral models \citep{konopacky2010,burgasser2012}; examining low mass companions of more massive stars with well determined ages, metallicity, etc such as the T dwarf companion to G3 star, HD 19467 \citep{crepp2015}; using PRV and Gaia data to constrain the stellar mass-radius relationship for metal-poor, low mass objects using  transiting, low mass halo stars discovered by K2 \citep{saylor2018}.

\subsection{Solar System Science}
A large wavelength coverage, high-dispersion infrared spectrograph is key to characterizing the molecular composition of planets and small bodies in the Solar system, including distant Kuiper Belt objects (KBOs) and comets. For the well-known Jovian and Saturnian moons, the diffraction limit of the adaptively corrected 10-30 meter telescopes allows spatially resolving surface features (e.g. volcanoes on Io or cryovolcanoes on Europa) and performing detailed remote molecular characterization \citep{Chanover2019}\footnotemark[2]. HISPEC/MODHIS will enable new science on the planets in our Solar System.  High resolution, AO-fed spectroscopy can shed light on the methane abundance and variability across the Martian surface, an area of great interest for astrobiology \citep{Wong2019}\footnotemark[2].  It can also be used to investigate cloud features on the ice giant worlds like Neptune and Uranus.

\subsection{Galactic and Extra-galactic Science}

HISPEC/MODHIS will be workhorse instruments for a wide range of Galactic science.  This includes abundance and kinematic characterization of individual stars in the field, in clusters and in other unique areas of the Milky Way, such as the Galactic Center \citep{Do2019}\footnotemark[2].  In the Galactic Center, there is also the opportunity to test fundamental physics such as the constancy of fundamental constants in the presence of an extreme gravitational field.  The spatial resolution of HISPEC/MODHIS plus the spectral resolution are required to look at variations of the fine structure constant, for example.  Additional effects of general relativity from the measurement of orbits of stars near the SgrA* will be more easily detected with high resolution spectroscopy than with continued astrometric monitoring, offering a unique niche for both HISPEC and MODHIS in this field.

HISPEC/MODHIS will also perform extragalactic science, including follow-ups of ultra-luminous infrared transients (e.g. supernovae in nearby galaxies) from ZTF and LSST, studies of stars in ultra-compact dwarf galaxies, and spectro-astrometry of rotating gas disks for the detection of supermassive black holes in galactic nuclei.  The instruments can be used to detect intermediate mass black holes, characterize AGN mergers, and investigate stellar feedback in nearby galaxies.  Both HISPEC and MODHIS are being designed with laser guide star AO systems, which opens up a wide range of faint object science previously unavailable to current instrumentation.



\section{Technical Overview}

In this section, we describe HISPEC, which is the diffraction-limited  high-resolution infrared spectrograph concept common to both Keck and the TMT-MODHIS facility. At Keck, HISPEC is to be fed by the front-end fiber injection unit of the Keck Planet Imager and Characterizer \citep[KPIC, see][]{Mawet2016,Mawet2018,delorme2018}. TMT-MODHIS includes the HISPEC spectrograph and a new front-end instrument interfacing the spectrograph to the first-light AO system of TMT, NFIRAOS \citep[Narrow Field InfraRed Adaptive Optics System,][]{Herriot2014}.  

\subsection{Front-end instrument}

The front-end instrument (FEI) is the essential link between the AO system and the single-mode spectrograph. Its purpose is to inject the diffraction-limited beam of the target into one or several SMFs and maintain accurate alignment throughout long-exposure observations. The pointing accuracy and stability is achieved through active sensing and control of the target and fiber positions using a scheme similar to \citet{Colavita1999}.

   \begin{figure*}
        \centering
        \includegraphics[width=0.8\textwidth]{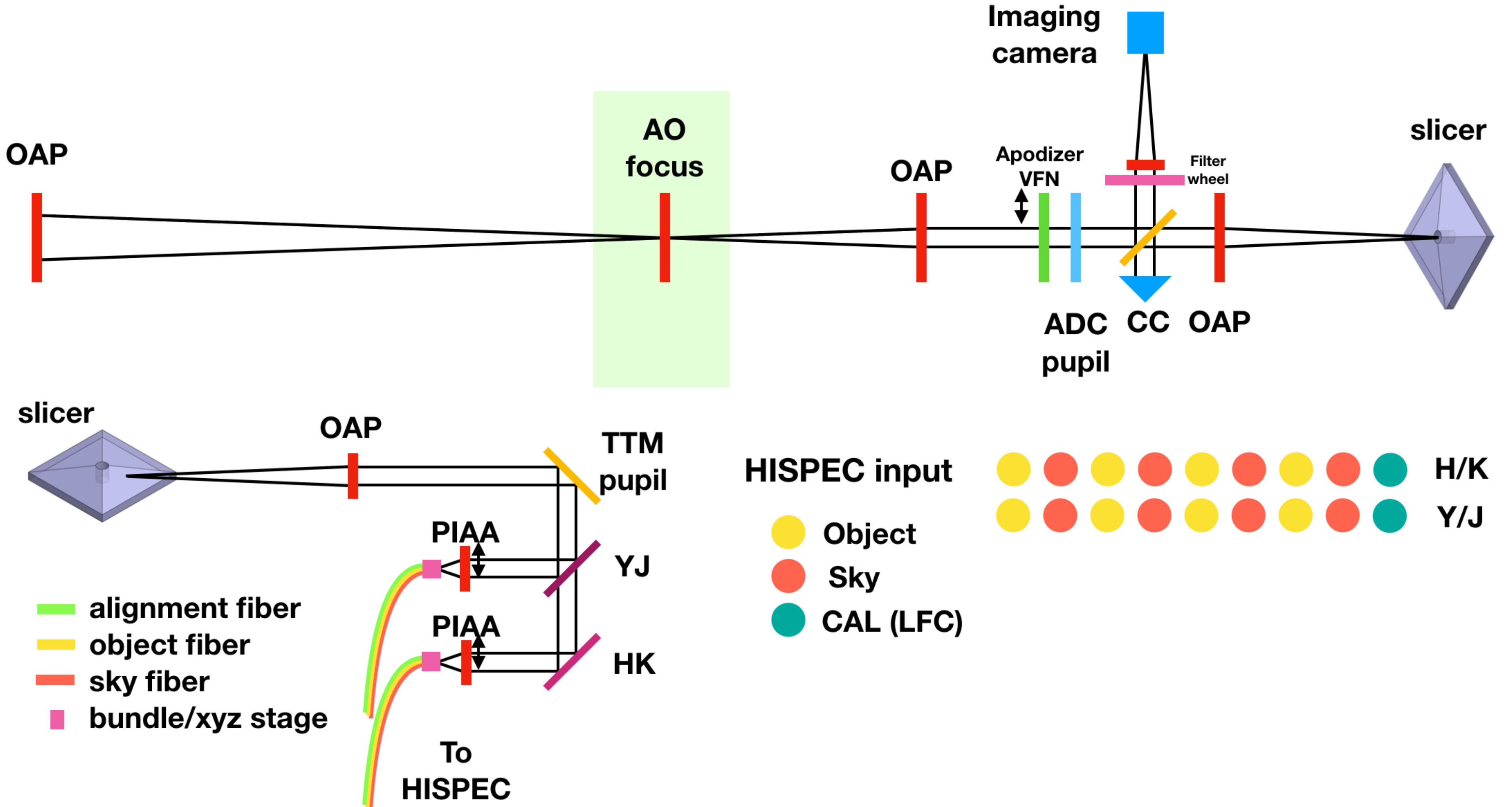}
        \caption{HISPEC/MODHIS Front-End Instrument (FEI) notional concept.}
        \label{fig:fei}
    \end{figure*}

An actuated tip-tilt mirror (TTM) is used to align the target image position with the tip of the SMF, whose relative locations are determined by simultaneously imaging the scene and the SMF on to a acquisition/tracking imaging camera (Figure \ref{fig:fei}). A beamsplitter (BS) or dichroic reflects part of the science beam to the tracking camera directly after the TTM. To locate the SMF, a light source is retro-fed through a set of reference alignment fibers located close to the science fiber. Ideally the reference fibers are part of the same bundle as the science fibers, ensuring mechanical stability of their relative positions. The BS reflects light from the SMF towards a corner cube (CC) retroflector, which sends the beam back through the BS and towards the tracking camera. A beacon image is formed on the tracking camera at the location of the SMF. The beacon is used to determine the TTM settings to co-align the object image and the SMF. 

The FEI is also designed to include high contrast capabilities for faint off-axis sources (e.g., exoplanets) using various pupil plane apodizers and masks, and provide feedback mechanisms for starlight suppression using the upstream AO systems (Keck AO or NFIRAOS on TMT). Optimized Phase Induced Amplitude Apodization (PIAA) lenses are used to remap the input beam from the Keck and TMT apertures into quasi-Gaussian beams matched to the fundamental mode of the SMF as in \citet{Jovanovic2017}, enabling close to ideal injection efficiency's in the diffraction limit ($\simeq$ 90\%).
 
Multiplexing can be ensured by using image slicing techniques, where the focal plane is divided in sectors, which can all be addressed with an independent fiber injection unit (Figure \ref{fig:fei}). Each fiber injection unit includes a dedicated TTM that can patrol the field of view on the fixed fiber bundle.

\subsection{Spectrograph}

\begin{figure*}[!t]
        \centering
        \includegraphics[width=0.6\textwidth,angle=-90]{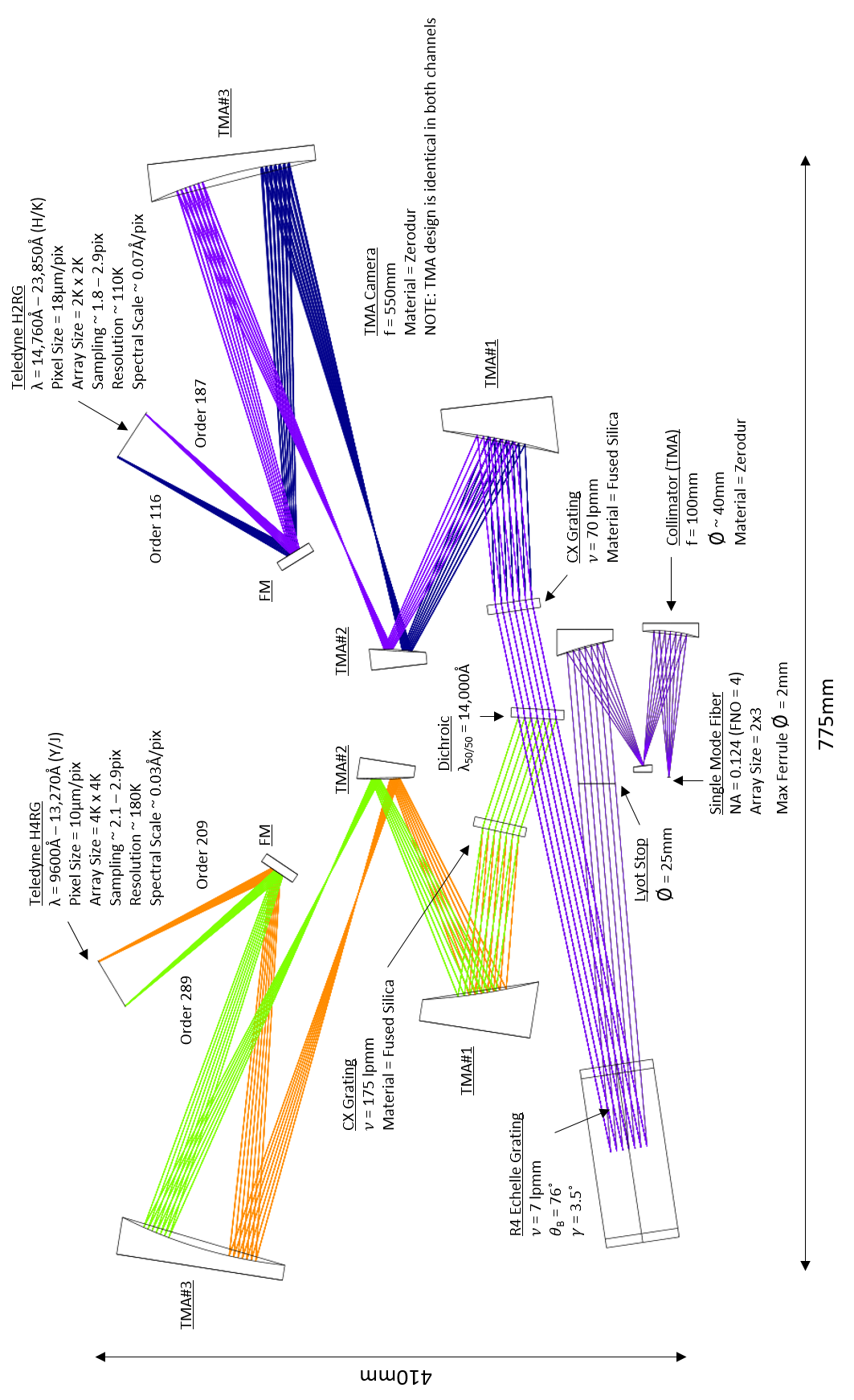}
        \caption{The HISPEC conceptual optical layout for the design incorporating a TMA F/4 collimator, a reflection grating ruled with 7 lines mm$^{-1}$, a VPH cross-disperser, and a TMA F/22 camera.  The spectrograph gives a fixed spectral format on the detector array in an overall compact design form.}
        \label{fig:OPTLAYOUT}
\end{figure*}

The HISPEC notional optical design (see Figure \ref{fig:OPTLAYOUT}) is a diffraction-limited echellette spectrometer with an almost all reflective design with a spectral resolving power of R$\sim$180,000 and R$\sim$110,000 in the yJ and HK passbands, respectively.  The full wavelength range from 9600 \AA\@ - 23,850 \AA\@ is broken up into two wavelength channels from 9600 \AA\@ - 13,270 \AA\@ (i.e. yJ-bands) and 14,760 \AA\@ - 23,850\AA\@ (i.e. HK-bands).  The wavelength gap from 13,270 \AA\@ - 14,760 \AA\@ is not covered by either channel, but is also where there is strong absorption from the atmosphere.  The main reasons to have two wavelength channels in the spectrograph are: light does not propagate as a single mode in a SM fiber for much more than a 1 octave spectral range and secondly to achieve close to Nyquist sampling over the full operating wavelength range where there is a 2.5$\times$ factor from the shortest to the longest wavelength.         

Light enters the spectrograph through a fiber array (2 columns). Each individual wavelength channel sees 1 column of several science, sky, and calibration fibers. The fiber array is collimated by an F/4 Three Mirror Anastigmat (TMA).  The collimator forms a circular pupil with a 25 mm diameter where a physical aperture stop is placed.

The relatively fast TMA can collimate a fiber ferrule diameter up to 2 mm in size and therefore is a optical design that is flexible enough to support any potential upgrades in wavelength coverage and fiber channel multiplicity. A single Off-Axis Parabola (OAP) collimator does not provide a sufficiently large enough diffraction-limited FOV at a F/4 focal ratio for the HISPEC fiber array and is limited to a field diameter size of $\sim$0.4 mm.  The TMA collimator design will also make centering the fiber array less sensitive to mis-alignments compared to an OAP collimator design.

The collimated light then gets dispersed in the spectral direction by reflecting off of a 7 lines per mm R4 echelle grating.  The grating is used in a quasi-Littrow condition where an out-of-plane angle $\gamma$ of 3.5$^\circ$ is needed to separate the incoming and outgoing light from the grating surface sufficiently enough to package the remaining optical elements without interference. The echelle grating splits the yJ-bands into 81 total spectral orders ranging from 209 - 289 and the HK-bands into 72 total spectral orders ranging from 116 - 187.   

The dispersed light from the grating is then separated into its two wavelength channels via a dichroic mirror.  The yJ-band is in reflection and the HK-band is in transmission, where the dichroic 50/50 transition edge is at 14,000 \AA\@.  The light then travels into the individual cameras after passing through a Volume Phase Holographic (VPH) cross-dispersing transmission grating.  The VPH cross-disperser allows for a simpler and almost symmetric layout.  To first order a VPH grating has a constant angular dispersion, which allows for a more uniform separation of the orders in the cross-dispersed direction and therefore allow for higher multiplicity of fibers to be re-imaged between the orders.  With a cross-dispersing prism the non-linear change of index of refraction with wavelength yields a closer spacing of the spectral orders containing the longest wavelengths therefore limiting the allowed fiber multiplexing in the spectrograph.  During the next design phase further study will have to be made about the cryogenic optical quality and stability of a VPH grating.  Otherwise cross-dispersing prisms will have to be used.  Most likely two ZnSe prisms with a refractive index of $\sim$2.4 will have to be used in the yJ channel and a single Silicon prism with a refractive index of $\sim$3.4 for the HK channel in order to get enough cross-dispersing power with reasonable prism apex angles to spread the spectral orders over the full height of the detector array.  The dichroic and cross-disperser are the only two refractive elements in the spectrograph, where changes in the index of refraction caused by thermal variations constrains any spectrum shifts at the detector to the cross-dispersed direction.

The individual cameras then re-images 1 column of fibers onto their respective detectors.  The HK channel will use a Teledyne H2RG 2.5$\mu$m cut-off detector with a 18$\mu$m per pixel size (upgrade option: H4RG 2.5$\mu$m cut-off detector), while the yJ channel will use a Teledyne H4RG with a 10$\mu$m per pixel size.  This roughly 2$\times$ factor in pixel size between the detector arrays means the two channels can utilize an identical camera design, while maintaining proper sampling.  This will help in lowering the cost of the cameras since they can be fabricated from the same parent optic.  The camera design is also a TMA with a focal ratio of F/22, which gives diffraction-limited image quality (i.e., $>$ 95\% Strehl) over the entire detector area in both channels.  The yJ channel has a spectral resolution of R$\sim$180,000 with a pixel sampling ranging from 2.1 - 2.9 pixels and a spectral sampling of $\sim$0.03 \AA/pix.  The HK channel has a spectral resolution of R$\sim$110,000 with a pixel sampling ranging from 1.8 - 2.9 pixels and a spectral sampling of $\sim$0.07 \AA/pix.  

\begin{figure}[!t]
        \centering
        \includegraphics[width=0.5\textwidth]{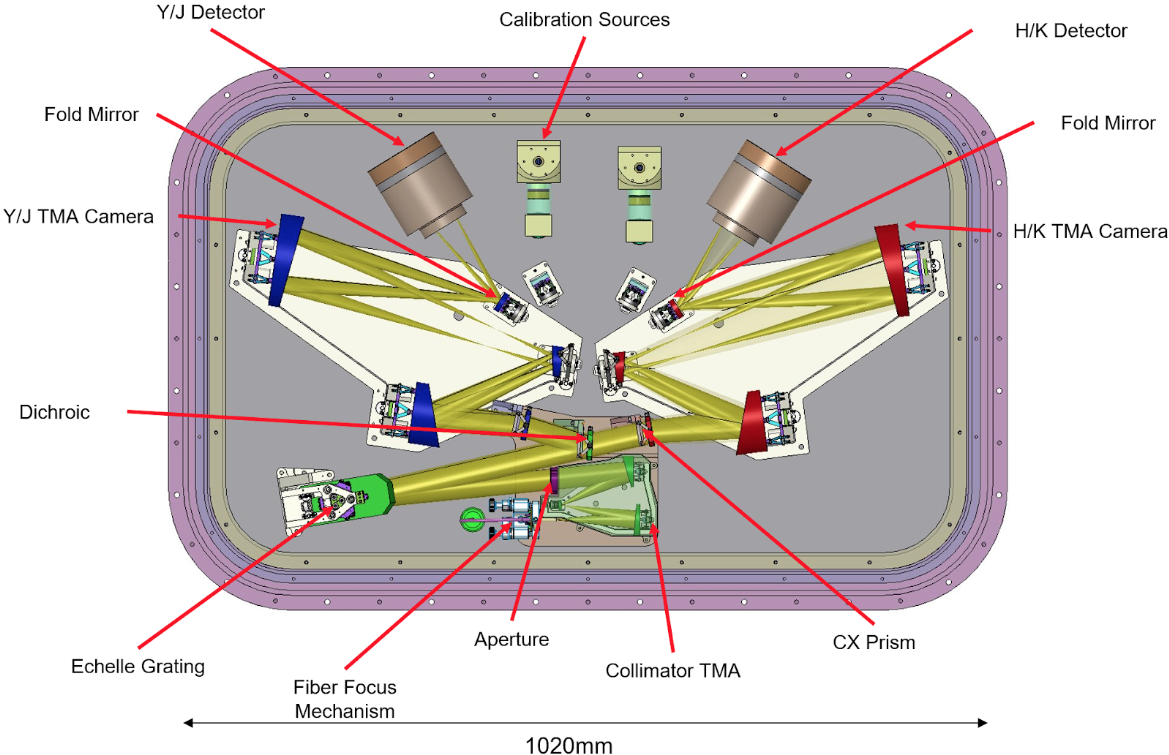}
        \caption{The HISPEC spectrograph conceptual mechanical layout.}
        \label{fig:omech}
\end{figure}

A notional mechanical design of the spectrograph is shown in Figure \ref{fig:omech}, featuring a compact and simple layout, optimized for thermal and mechanical stability.

\section{Technology Drivers}
Although the front-end instrument is based on mostly mature technologies, there are some aspects that will require further development to determine the most efficient and robust way to realize the instrument in time for TMT first light.

\textbf{Beam shaping optics:} the PIAA lenses have been successfully used in the past \citep{Jovanovic2017} but require further development to optimize them for the two discrete wavebands of the HISPEC spectrograph (yJ and HK).

\textbf{Image slicing optics:} the choice of how to slice the image will be partially driven by the layout of science fields to be observed and partially by the types and quality of optics available. Further development may be needed here. 

\textbf{Optimal tracking strategies:} there are very few SMF injections operating on-sky. Maintaining alignment between the science target and the SMF to within $\approx 0.1\lambda/D$ will be required to maintain efficient coupling. Optimal strategies to provide this level of alignment are being proof-tested by KPIC. 

\subsection{Low loss infrared single-mode fibers}
HISPEC/MODHIS will exploit SMFs because their output beam profile does not evolve with time and is independent of the input illumination profile. This will provide superior stability for the line spread function (LSF) for Doppler applications. In addition, as outlined above, a diffraction-limited feed decouples the spectrograph performance from that of the telescopes aperture allowing for a highly compact yet very high resolution instrument to be realized, substantially reducing footprint, mechanical deflection and cost. It also means that a single spectrograph could be built and used at both Keck then TMT.

However, SMFs have a limited range of wavelengths over which they are  single-mode. There are no SMFs covering the full y-K wavelength range. Therefore the FIU will split the light in two passbands and inject the light into two separate fiber bundles that each span narrower wavelength ranges (yJ and HK). There are many suitable options yJ SMFs, including OFS BF05635-02 fiber (core and cladding diameters of 4.4 $\mu$m and 125 $\mu$m, a Numerical Aperture (NA) of 0.16, and a single mode cut-off wavelength of 9600 \AA\@). For the HK SMF, one option is offered by Le Verre Fluore (LVF), who produce some of the highest performance fluoride fibers. The best fit stock fiber is \#3051 (core  and cladding diameters of 7.0 $\mu$m and 125 $\mu$m, a NA of 0.17, and a single mode cut-off wavelength of 15,500 \AA\@), which is not single-mode across the entire H-band. A custom fiber will need to be developed with LVF that operates across HK with minimal losses. Alternatively, one could consider shifting the split from between the J and H bands to in the middle of the H band, but then the location of the dichroic edge would need to be carefully considered to not coincide with important spectral features. 

\subsection{Few mode fibers and photonic lanterns} 
Although the priority is to utilize SMFs, it may be favorable to use fibers with higher collection efficiency's in the shorter wavebands (yJ) where the Strehl ratio and hence coupling efficiency will be lower. For this there are two options: few mode fibers and photonic lanterns. Few mode fibers lay at the boundary between multimode fibers (MMF) and SMFs and support as their name implies, a few modes (typically 2-8). By supporting more modes these fibers typically allow for higher coupling efficiency than a SMF~\citep{Horton2007}. Owing to the fact that they support several modes, ensuring the output beam is homogenized and hence stable with time will require active scrambling. Alternatively, a photonic lantern consists of a MMF at the input end and series of SMFs at the other end. The light makes a gradual transition from the MMF end, which is trivial to couple into to a series of SMFs which offer diffraction limited performance to the spectrograph~\citep{sergio2013,birks2015}. In this way the flux will be split amongst fibers reducing the flux/channel, but the total flux will be much higher than for a SMF, especially at shorter wavelengths where the AS performance is reducing. We will source commercially available versions of both fiber types and conduct laboratory experiments to assess their applicability for HISPEC/MODHIS. We will then work with vendors and other research institutes to customize these devices for our application. This development is expected to occur over the next 5 years.  

\subsection{Infrared Laser Frequency Combs}
Infrared laser frequency combs (LFCs) will provide long term,  sub-m/s precision across the HISPEC band. Devices now in hand or in development use Electro-Opical Modulation (EOM) to generate combs with intrinsic spacings of 10-30 GHZ which are readily resolvable with R$\sim100,000$ spectrographs without the need for extensive pre-filtering. These combs have been demonstrated at Keck~\citep{yi2016} and with HPF at the Hobby-Eberly telescope~\citep{metcalf2019}. They can either be referenced to a stable laser line such as acetylene ($<$ 30 cm/s), to a second, 100 MHz comb ($<$10 cm/s), or via $f-2f$ self-referencing (sub cm/s stability tied to an SI frequency standard). New LFC technologies based on micro-resonator microcombs promise to greatly reduce the size, cost and complexity of NIR LFCs with a prototype already demonstrated at Keck~\citep{suh2019}.  Future LFC's will be developed with extended wavelength range to cover y-K. We plan to further develop this new technology with significant investment over the next 5-7 years. 

\section{Organization, Partnerships, and Current Status}

HISPEC/MODHIS is currently led by Caltech, UC Los Angeles, and UC San Diego, with science team members from the entire US community, and TMT International Observatory (see author list). Participation in the instrument construction is currently being discussed with partners from Canada (NRC, University of Montreal) and Japan. We expect to extend the instrument team to TMT international partners, including China and India.

\section{Schedule and Cost Estimates}
In this section, we focus on the schedule and cost estimate for the TMT-MODHIS facility, which includes the full cost of the HISPEC spectrograph. As mentioned before, the fabrication and deployment of HISPEC is currently being fast-tracked for Keck Observatory as a pathfinder to TMT-MODHIS.

\subsection{Schedule}

\begin{table}[!t]
\begin{tabular}{c|c|c|c}
\textbf{Phase}                                                        & \textbf{\begin{tabular}[c]{@{}c@{}}Cal.\\ Years\end{tabular}} & \textbf{\begin{tabular}[c]{@{}c@{}}Labor\\ (wk yrs)\end{tabular}} & \textbf{\begin{tabular}[c]{@{}c@{}}Total\\ Cost\end{tabular}} \\ \hline
\textbf{\begin{tabular}[c]{@{}c@{}}Conceptual\\ Design\end{tabular}}  & 1.00                                                              & 6.1                                                                 & \$1.714M                                                      \\ \hline
\textbf{\begin{tabular}[c]{@{}c@{}}Preliminary\\ Design\end{tabular}} & 1.00                                                              & 8.5                                                                 & \$3.190M                                                      \\ \hline
\textbf{\begin{tabular}[c]{@{}c@{}}Final\\ Design\end{tabular}}       & 1.25                                                              & 11.0                                                                & \$3.265M                                                      \\ \hline
\textbf{Fabrication}                                                  & 1.50                                                              & 9.2                                                                 & \$10.311M                                                      \\ \hline
\textbf{Integration}                                                  & 1.25                                                              & 16.5                                                                & \$4.474M                                                      \\ \hline
\textbf{AIV}                                                          & 1.00                                                              & 6.9                                                                 & \$1.854M                                                      \\ \hline \hline
\textbf{Total}                                                        & 7.00                                                              & 58.2                                                                & \$24.808M                                                    
\end{tabular}
\caption{Development schedule for MODHIS, broken down by phase, duration, and labor.}
\label{table:schedule}
\end{table}

The MODHIS schedule and costing draws credibility from similarly scoped high-resolution spectrographs (e.g., PARVI), FEI (e.g., KPIC), and the other TMT first light instruments (IRIS and WFOS). We also expect to draw upon ongoing development of Keck-HISPEC to help accelerate the design phases of MODHIS. The development of MODHIS follows a 7 year schedule outlined in Table~\ref{table:schedule}. We anticipate the three design phases (conceptual, preliminary, and final) to last 2.25 years, followed by a 1.5 year fabrication phase, a 1.25 year integration phase, and finally a 1 year assembly, integration, and validation (AIV) phase at the telescope.

\subsection{Cost Estimates}

The development of MODHIS is costed at just under \$25M, placing it at the low-end of the medium cost category for ground-based projects evaluated by ASTRO2020: ``Funding for this effort is expected to come from TMT and its Members''. A full cost estimate of developing MODHIS, including sufficient contingency, was performed in June 2019 by Caltech and TMT. Table~\ref{table:cost} itemizes MODHIS costs by work breakdown structure tasks -- the methodology is consistent with other TMT first light instruments. We highlight specific deliverables below. 

The ``Structure \& OIWFS" element contains the MODHIS interface and support structure that enables attachment to NFIRAOS (TMT first light AO system), the MODHIS front-end instrument rotator, and the on-instrument wavefront sensor (OIWFS) assembly, including the OIWFS detector cryostat and controllers.

The ``Front End Instrument" element includes the subsystem that acquires, directs, collects, and maintains alignment of the light at the output of NFIRAOS, delivering the light into the single mode fiber feeding the spectrograph. The cost includes all elements plotted in Figure~\ref{fig:fei}.

The costing of the ``Spectrograph" is directly informed by a detailed cost analysis of both PARVI and Keck-HISPEC. This subsystem includes the spectrograph cryostat and optics, two infrared H4RG detectors and associated readout electronics, the internal calibration system (including two stable Laser Frequency Combs), and all equipment needed for integration and testing, including summit AIV. 

The costing of the ``Instrument Control Software" element contains all software related activities required to support observing and testing, including the instrument sequencer, hardware control modules, interfaces to NFIRAOS, and hardware necessary for the development and prototyping of the software deliverables. 

The ``Integration \& AIV" element includes the integration of the MODHIS subsystems to form the overall MODHIS science instrument, including integrating the OIWFS with the front end instrument, the front end instrument with the fiber system, and the fiber system with the spectrograph. All of these integration steps will involve execution of progressive acceptance test plans culminating in a pre-ship readiness review. AIV includes hardware integration with NFIRAOS and the telescope, along with integrating the control software and data reduction pipeline to the observatory software environment.

\begin{table}[]
\begin{tabular}{c|c|c|c}
\hline
\textbf{Task}                 & \textbf{\begin{tabular}[c]{@{}c@{}}Labor\\ (wk yrs)\end{tabular}} & \textbf{\begin{tabular}[c]{@{}c@{}}Total\\ Cost\\ (\$)\end{tabular}} & \textbf{\begin{tabular}[c]{@{}c@{}}Total\\ Cost\\ (\%)\end{tabular}} \\ \hline
\textbf{\begin{tabular}[c]{@{}c@{}}Project\\ Management\end{tabular}}      & 3.7                                                                 & \$1.908M                                                           & 8                                                                  \\ \hline
\textbf{\begin{tabular}[c]{@{}c@{}}Systems\\ Engineering\end{tabular}}     & 6.2                                                                 & \$2.197M                                                           & 9                                                                  \\ \hline
\textbf{\begin{tabular}[c]{@{}c@{}}Science\\ Team\end{tabular}}            & 5.5                                                                 & \$1.065M                                                           & 4                                                                  \\ \hline
\textbf{\begin{tabular}[c]{@{}c@{}}Structure \&\\ OIWFS\end{tabular}}      & 11.1                                                                & \$4.875M                                                           & 20                                                                 \\ \hline
\textbf{\begin{tabular}[c]{@{}c@{}}Front End\\ Instrument\end{tabular}}    & 7.4                                                                 & \$2.924M                                                           & 12                                                                 \\ \hline
\textbf{\begin{tabular}[c]{@{}c@{}}Fiber\\ Management\end{tabular}}        & 1.8                                                                 & \$0.586M                                                           & 2                                                                  \\ \hline
\textbf{Spectrograph}                                                      & 8.1                                                                 & \$7.347M                                                           & 29                                                                 \\ \hline
\textbf{\begin{tabular}[c]{@{}c@{}}Instrument\\ Control SW\end{tabular}}   & 5.4                                                                 & \$1.383M                                                           & 6                                                                  \\ \hline
\textbf{\begin{tabular}[c]{@{}c@{}}Data Reduction\\ Pipeline\end{tabular}} & 1.5                                                                 & \$0.389M                                                           & 2                                                                  \\ \hline
\textbf{\begin{tabular}[c]{@{}c@{}}Integration\\ \& AIV\end{tabular}}      & 7.5                                                                 & \$2.134M                                                           & 8                                                                  \\ \hline \hline
\textbf{Total}                                                             & 58.2                                                                & \$24.808M                                                          & 100
\end{tabular}
\caption{Development cost for MODHIS broken down by task, labor, and cost.}
\label{table:cost}
\end{table}


\clearpage

{\footnotesize
\bibliography{wp}}


\end{document}